%% file: snfract1A.tex
\documentclass[prb,superscriptaddress,showpacs,showkeys,twocolumn]{revtex4-1}
\usepackage{amsfonts,bbold,amsmath,amssymb,graphicx,epstopdf,verbatim,dsfont,color}
\usepackage[english]{babel}

\begin{document}

\title{Normal and superfluid fractions of inhomogeneous nonequilibrium quantum fluids}
\author{Vladimir N. Gladilin}
\affiliation{TQC, Universiteit Antwerpen, Universiteitsplein 1,
B-2610 Antwerpen, Belgium}
\author{Michiel Wouters}
\affiliation{TQC, Universiteit Antwerpen, Universiteitsplein 1,
B-2610 Antwerpen, Belgium}
\date{\today}

\begin{abstract}

We present a theoretical analysis of the normal and superfluid fractions of quantum fluids described by a nonequilibrium extension of the Gross-Pitaevskii equation in the presence of an external potential. Both disordered and regular potentials are considered. The normal and superfluid fractions are defined by the response of the nonequilibrium quantum fluid to a vector potential, in analogy with the equilibrium case. We find that the physical meaning of these definitions breaks down out of equilibrium. The normal and superfluid fractions no longer add up to one and for some types of external potentials, they can even become negative. 
\end{abstract}
\maketitle

\section{Introduction}

The hydrodynamic behavior of multimode laser systems has been
recognized long ago~\cite{brambilla1991,staliunas1993}, but it is on
the platform of semiconductor exciton-polaritons that a systematic
investigation of the hydrodynamics of photonic systems has
been performed \cite{iac_review}. Microcavity polaritons are hybrid
light matter excitations in semiconductor heterostructures
consisting of a quantum well embedded inside an optical microcavity
\cite{deveaud2007}. When the optical mode is resonant with the
quantum well exciton frequency, the quasi-particles are the coherent
superpositions of exciton and photon. Thanks to the photonic
component, the microcavity polariton is much lighter than the bare
exciton, which enhances the coherence properties. Thanks to the
excitonic component, polaritons show a significant interaction
strength. Being bosonic particles, microcavity polaritons can show
long range phase coherence in the quantum degenerate regime, giving
rise to superfluid \cite{leggett1999} properties, such as quantized
vortices \cite{lagoudakis2008}.

The photonic component also introduces losses, which makes it so far
impossible to achieve fully equilibrated Bose-Einstein condensation.
In order to compensate for the finite polariton life time, particles
have to be injected. The two main schemes are resonant and
nonresonant excitation. In the former regime, the velocity of the
polariton fluid can be tuned by varying the angle of the excitation
laser. With this pumping scheme, superfluidity according to the
Landau criterion (absence of scattering off defects) \cite{amo2009}
and the nucleation of quantized vortices \cite{nardin2011} as well
as solitons \cite{amo2011} have been experimentally observed.
Despite the finite polariton life time, the observed phenomenology
in those experiments was close to the equilibrium behavior.

From a conceptual viewpoint, the main disadvantage of resonant
excitation is the fact that the pumping laser fixes the phase of the
quantum fluid, which excludes the study of superfluidity in the
sense of linear response or metastable superflow. It has been
argued~\cite{iac_review} that the most precise and quantitative
definition of superfluid $f_s$ and normal $f_n$ fractions involves
the response of the quantum gas to a weak transverse vector
potential ${\bf A}$ as analyzed in Ref.~\onlinecite{hohenberg65}.
For a spatially homogeneous system of density $n$, the normal
fraction $f_n$ can be defined as
\begin{eqnarray}
f_n=\lim_{q\to 0} \frac{m}{n}\chi_T({\bf q}). \label{fn}
\end{eqnarray}
Here $m$ is the effective mass and  $\chi_T({\bf q})$ is the
susceptibility tensor relating the average transverse current to the applied
vector field in Fourier space.

The major question is here to what extent the nonequilibrium
character of polariton condensates affects their superfluid
properties. A first theoretical study was performed by Keeling on
the superfluid fraction of a homogeneous driven-dissipative
polariton condensate with a Keldysh functional integral approach
\cite{keeling2011}. Out of equilibrium, a nonvanishing normal
component is always present, due to nonequilibrium fluctuations.

The work by Janot et al.~\cite{janot13} addressed superfluidity in
the disordered case. A reduction in superfluid fraction due to
disorder is due to the pinning of the fluid in the potential minima
and a first theoretical estimate can be obtained at the mean field
level~\cite{lieb2002,fontanesi2010}. In the theoretical study of
Ref.~\onlinecite{janot13}, the response was studied to a small twist
$\theta$ of the phase  between two boundaries of the condensate
separated by its size $L$. This is equivalent to the presence of a
weak vector potential ${\bf A}={\bf e}_x \theta/L$. As argued
in Ref.~\onlinecite{janot13}, the superfluid stiffness in this case
can be estimated in accordance with the equilibrium definitions~\cite{leggett70,fisher73} as
\begin{eqnarray}
f_s= \frac{\omega(A)-\omega(0)}{A^2},\label{fs}
\end{eqnarray}
where $\hbar \omega$ is the chemical potential of the condensate.

The weak vector potential ${\bf A}={\bf e}_x \theta/L$ corresponds to a slow rotation at a velocity of less than one quantum of circulation $v<2\pi/L$. Due to the phase quantization, the superfluid cannot rotate, so that the current is supported by the normal phase only (Hess-Fairbank effect) \cite{leggett1999}. This allows us to define the normal fraction in terms of the dependence of the current density on the gauge field
\begin{eqnarray}
f_n= \frac{\langle j_x(A_x)\rangle
}{nA_x}. \label{fn2a}
\end{eqnarray}

The application of (synthetic) gauge fields in optical systems has
developed into an active field of research~\cite{lu2014}, thanks to
several experimental
breakthroughs~\cite{hafezi2011,rechtsman2013,rechtsman2013b} and a
wealth of theoretical proposals. Seen from this context, we address
the question of the effect of a small gauge field on a the frequency and current in a nonequilibrium condensate.

\section{Model}
We will apply the aforedescribed definitions of $f_s$ and $f_n$
 to the systems described by the generalized Gross-Pitaevskii
equation (gGPE)~\cite{prl2007,keeling2008}, which is taken in the
form~\cite{iac_review}
\begin{eqnarray}
i\hbar \frac{\partial \psi}{\partial t} =&&
\left[-\frac{\hbar^2\nabla^2}{2m} +g |\psi|^2 + V\right. \nonumber
\\
&&\left.+\frac{i}{2} \left(\frac{P}{1+|\psi|^2/n_s}-\gamma \right)
\right] \psi,
\label{eq:ggpe}
\end{eqnarray}
with a static potential $V({\bf r})$  and a contact interaction with
the strength $g$. The imaginary term in the square brackets on the
right hand side describes the saturable pumping (with strength $P$
and saturation density $n_s$), that compensates for the losses
($\gamma$). The physical origin of the pumping term for
exciton-polariton condensates is an excitonic reservoir that is
excited by a nonresonant laser.

In general, the pumping intensity $P$ is coordinate dependent and
can be represented as $P({\bf r})=P_0p({\bf r})$, with
$P_0=\max{P({\bf r})}$. Assuming that the interaction strength is
positive, $g>0$, it is convenient to rewrite Eq.~(\ref{eq:ggpe}) in
a dimensionless form, by expressing the particle density $|\psi|^2$
in units of $n_0\equiv n_s(P_0/\gamma-1 )$, time in units of
$\hbar/(gn_0)$, and length in units of $\hbar/\sqrt{2mgn_0}$:
\begin{eqnarray}
i\frac{\partial\psi}{\partial t}=&& \left[-\nabla^2 +|\psi|^2 + V_0v
\phantom{\frac{\psi^2}{\psi^2}}\right. \nonumber
\\
&&\left.+ic\frac{1-|\psi|^2+(1+\nu^{-1})(p-1)}{1+\nu |\psi|^2}
\right] \psi . \label{eq:ggpe2}
\end{eqnarray}
Equation~(\ref{eq:ggpe2}) contains three dimensionless scalar
parameters: $V_0=\max{|V({\bf r})|}/{gn_0}$, $c=\gamma/(2gn_s)$ and
$\nu=n_0/n_s$. The dimensionless functions $p({\bf r})$ and $v({\bf
r})=V({\bf r})/(V_0gn_0)$ describe the spatial distributions of the
pumping intensity and static potential, respectively.

Our numerical simulations are performed for a region of sizes
$L_x\times L_y$ with periodic boundary conditions in the $x$
direction and the Neumann boundary conditions at $y=0, L_y$. A
uniform grid with $N_x\times N_y$ nodes is used. A vector potential,
chosen as ${\bf A}=(A_x, 0,0)$ with constant $A_x$, is introduced by
replacing $\partial/\partial x$ with $\partial/\partial x -iA_x$ in
Eq.~(\ref{eq:ggpe2}). For electrically charged particles, our
configuration would correspond to a 2D cylindrical shell in an
axially symmetric magnetic field parallel to the cylinder axis. 
We should keep in mind that
in the case of an inhomogeneous static potential and/or pumping
intensity, described by the functions $v({\bf r})$ and $p({\bf r})$,
respectively, the average current $\langle j_x\rangle$ can be
nonzero even at $A=0$. A natural generalization of the
expression \eqref{fn2a} for $f_n$ to this case seems to be
\begin{eqnarray}
f_n= \frac{\langle j_x(A_x)\rangle -\langle j_x(0)\rangle
}{nA_x},\label{fn2}
\end{eqnarray}
where the current density in the units used is given by the
expression
\begin{eqnarray}
{\bf j}= {\rm Im}\left[\psi^*\nabla \psi\right].\label{j}
\end{eqnarray}

\section{Random potentials}

Irregular potential landscapes are highly relevant for experimental
realisations of polariton condensation because of growth
imperfections in the semiconductor heterostructures. At equilibrium,
the interplay between Bose-Einstein condensation and disorder gives
rise to rich physics, with a zero temperature superfluid to
Bose-glass quantum phase transition\cite{fisher1989}.

\begin{figure}[h!]
\centering
\includegraphics*[width=1\linewidth]{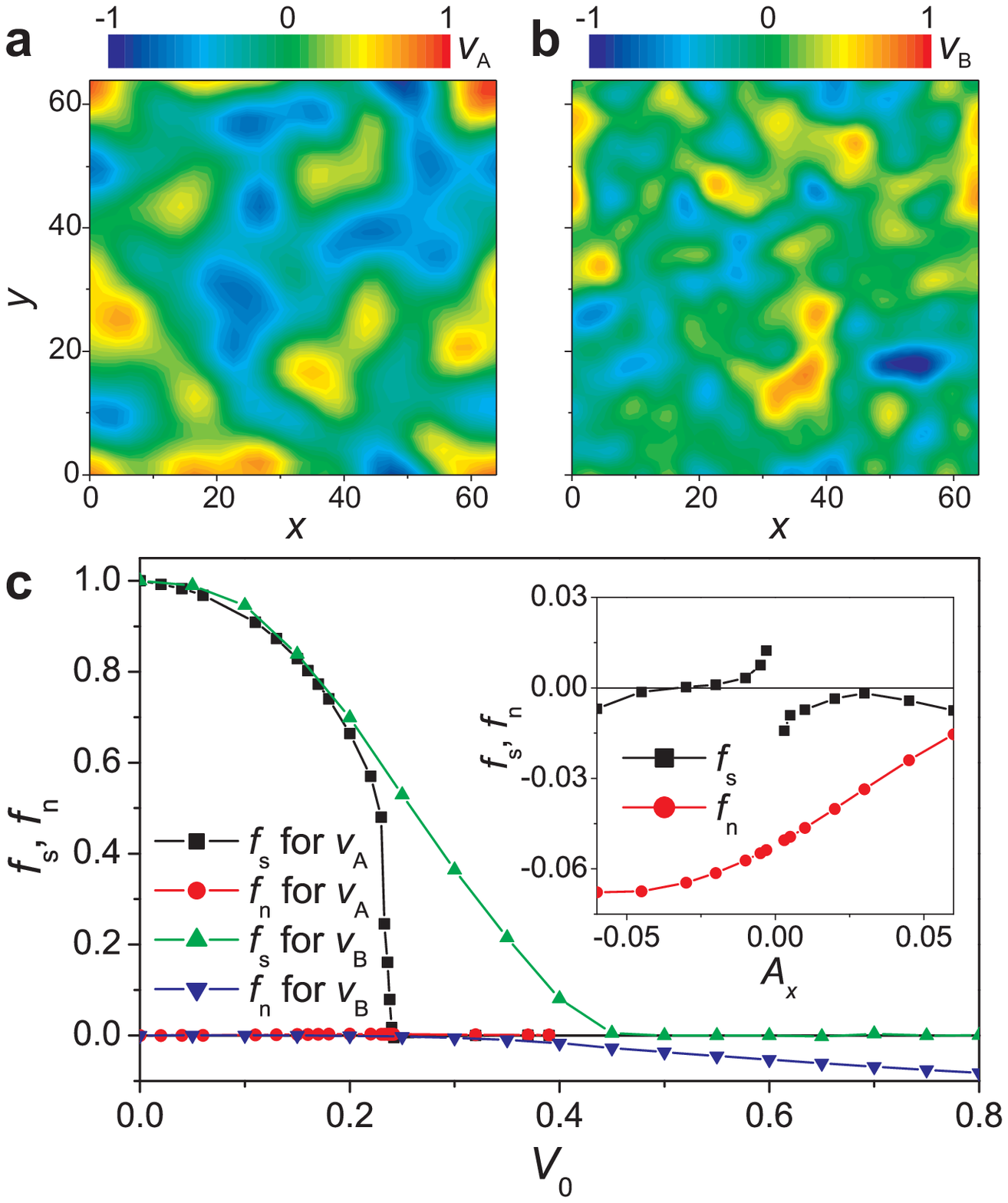}
\caption{Distributions $v_A(x,y)$ (panel a) and $v_B(x,y)$ (panel b)
for two random potentials with the correlation length $\xi_v = 5$
and 3, respectively. (c) $f_s$ and $f_n$, calculated for $v=v_A$,
$L_x = L_y = 64$, $N_x = N_y = 32$, $A_x = 0.1$ and for $v=v_B$,
$L_x = L_y = 64$, $N_x = N_y = 64$, $A_x = 0.03$, as a function of
$V_0$. Inset: $f_s$ and $f_n$, calculated for $v=v_A$, $L_x = L_y =
64$, $N_x = N_y = 32$, $V_0 = 0.24$, as a function of $A_x$. The
calculations are performed for $\nu =1$, $c = 1$, $p({\bf r}) = 1$.}
\label{Fig1}
\end{figure}

We consider the case of a uniform pumping intensity ($p({\bf r})=1$)
and a static potential, described by a random distribution $v({\bf
r})$. Two examples of the used random distributions, $v_A(x,y)$ and
$v_B(x,y)$, with the correlation length $\xi_v=5$ and 3,
respectively, are shown in Figs.~\ref{Fig1}a and \ref{Fig1}b.
In Fig.~\ref{Fig1}c we plot the quantities $f_s$ and $f_n$,
corresponding to these potential distributions, as a function of the
strength $V_0$ of the potential. In a qualitative agreement with the
results of Ref.~\onlinecite{janot13}, for both random potentials the
function $f_s$ decreases with $V_0$ and falls to zero at a
sufficiently strong potential. At the same time, in the whole range
of $V_0$ under consideration, the calculated $f_n$ remains close to
zero or even becomes negative at large $V_0$, so that the expected
``sum rule'' $f_s+f_n=1$ is obviously violated. At first sight, this
violation could be related exclusively to the calculated $f_n$
(e.g., to lack of physical meaning of expression~\eqref{fn2} or a
lack of numerical accuracy in the corresponding estimation).
However, a deeper analysis shows that the problem actually has a
more general character. Indeed, from the inset to Fig.~\ref{Fig1}c
one can see not only a pronounced asymmetry of $f_n(A_x)$, which
further illustrates inconsistencies in treating the found $f_n$ as
the normal fraction of the condensate, but also the presence of a
clear discontinuity of $f_s(A_x)$ at $|A_x|\to 0$. The latter allows
us to put under question also the possibility to interpret the
calculated $f_s$ as the superfluid stiffness. In other words, in the
presence of a random potential, neither Eq.~(\ref{fs}) nor
Eq.~(\ref{fn2}) seem to provide correct estimates for the superfluid
and normal fractions of the condensate described by
Eq.~(\ref{eq:ggpe2}). This fact can be attributed to the existence
of non-negligible in-plane currents in the condensates under
consideration, even at $A=0$. Below we will illustrate the above
statement by few simple examples with regular potentials.

\section{Regular potentials}

\begin{figure}[h!]
\centering
\includegraphics*[width=1\linewidth]{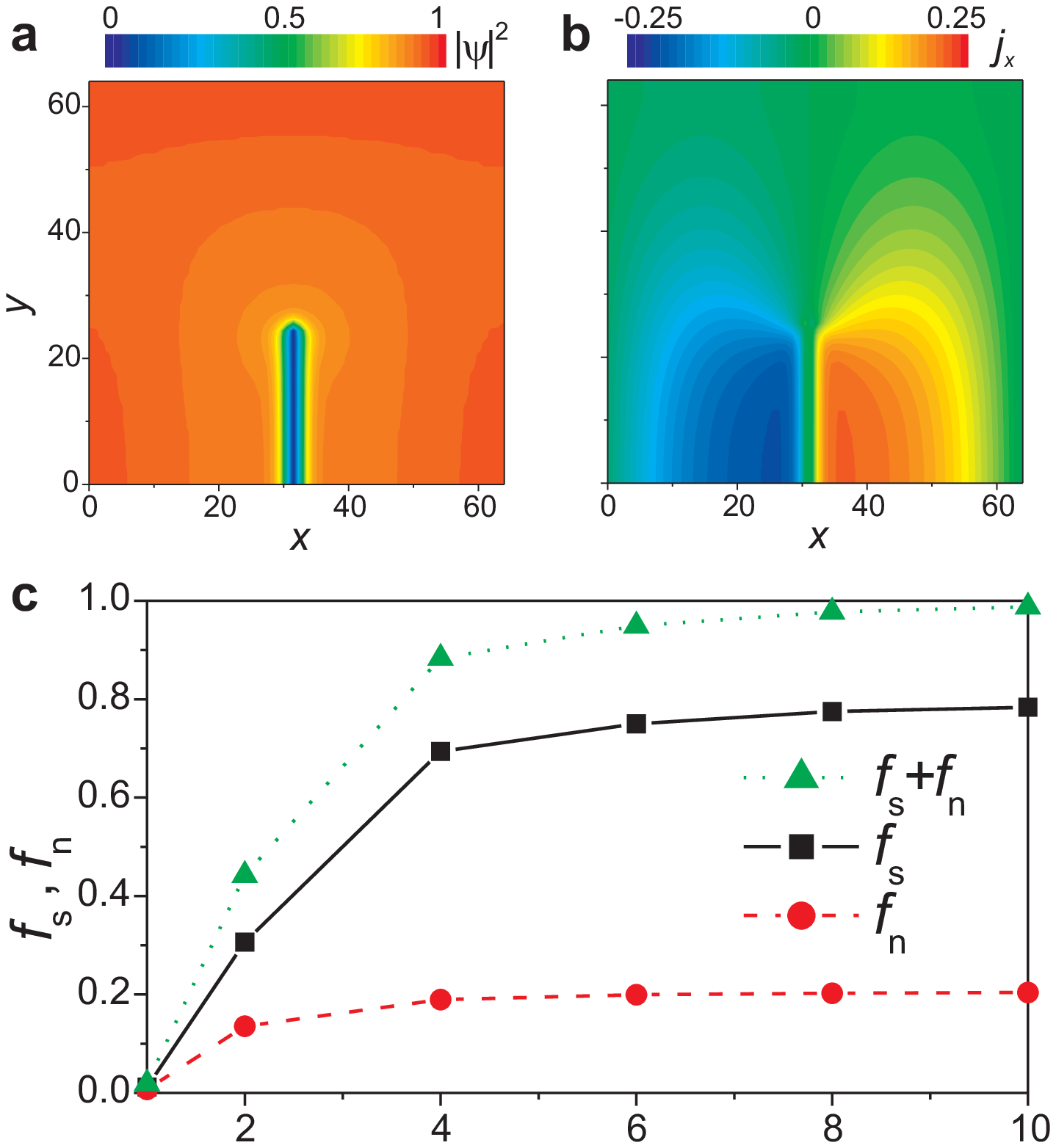}
\caption{Density $|\psi|^2$ (panel a) and current $j_x$ (panel b)
distributions, calculated for a system with  $L_x = L_y = 64$, $N_x
= N_y = 32$, $L_{\rm cut}=0.4 L_y$, $V_0=0$, $p=1$ and $c=1$  at
$\nu=2$, $A_x = 0$. (c) $f_s$ and $f_n$, calculated for the same
system at $A_x = 0.05$, as a function of $\nu$. } \label{Fig2}
\end{figure}

First, let us consider a structure with a partial cut in the
direction perpendicular to the vector potential ${\bf e}_xA_x$. The
cut is introduced through the additional condition $\left.
\psi\right|_{x=L_x/2,y\leq L_{\rm cut}}=0$. An example of the
density and current distributions in a system with such a cut, as
given in Figs.~\ref{Fig2}a and \ref{Fig2}b, respectively,
corresponds to $V_0=0$, $p=1$, $c=1$ and $\nu =2$.
At those parameter values the contribution of the pumping-loss term
to gGPE is relatively large [see Eq.~(\ref{eq:ggpe2})] leading to
the appearance of rather strong currents on both sides of the cut
(see Fig.~\ref{Fig2}b). The corresponding values of $f_s$ and $f_n$
as well as their sum are significantly smaller than one (see
Fig.~\ref{Fig2}c for $\nu\leq 2$). With increasing $\nu$, the role
of the pumping-loss term is suppressed, equation~(\ref{eq:ggpe2})
approaches the standard Gross-Pitaevskii equation, and the
quantities $f_s$ and $f_n$ gradually restore their physical meaning,
so that the sum $f_s+f_n$ tends to reach 1 at $\nu \sim 10$.

\begin{figure}[h!]
\centering
\includegraphics*[width=1\linewidth]{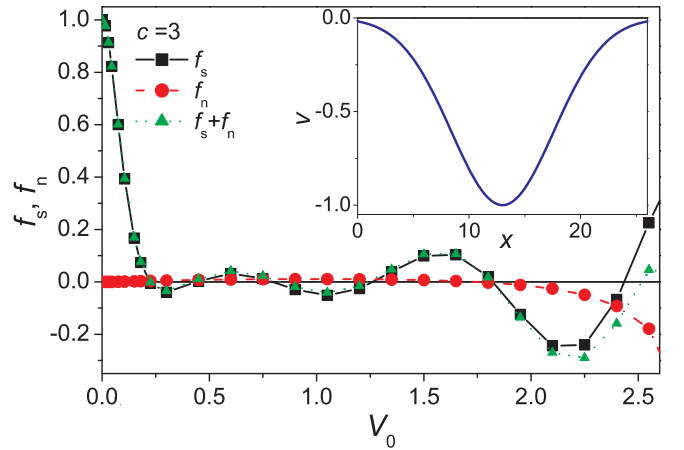}
\caption{$f_s$ and $f_n$ as a function of $V_0$  for  $L_x = 26$,
$N_x = 128$, $c = 3$, $\nu =2$, $p = 1$, $A_x = 0.04$. Inset: shape
$v(x)$ of the static potential (solid line) and density distribution
$|\psi(x)|^2$ at $V_0=0.45$ (dashed line). } \label{Fig3}
\end{figure}

In Fig.~\ref{Fig3} we show $f_s(V_0)$ and $f_n(V_0)$ calculated for
a system, where the aforementioned cut is replaced with a potential
well $V_0v(x)$, uniform in the $y$ direction (see the inset to
Fig.~\ref{Fig3}).
Figure~\ref{Fig3} corresponds to a relatively large value of the
parameter $c$ ($c=3$). In this case, the behaviour of $f_n(V_0)$
closely resembles that in the presence of a random static potential
(Fig.~\ref{Fig1}c). The quantity $f_s(V_0)$, shown in
Fig.~\ref{Fig3}, first manifests a decrease, qualitatively similar
to that in Fig.~\ref{Fig1}c. At larger $V_0$, this decrease is
overridden by oscillations of $f_s(V_0)$. Here, the definition
\eqref{fn} clearly loses its physical meaning of normal fraction.

\begin{figure}[h!]
\centering
\includegraphics*[width=1\linewidth]{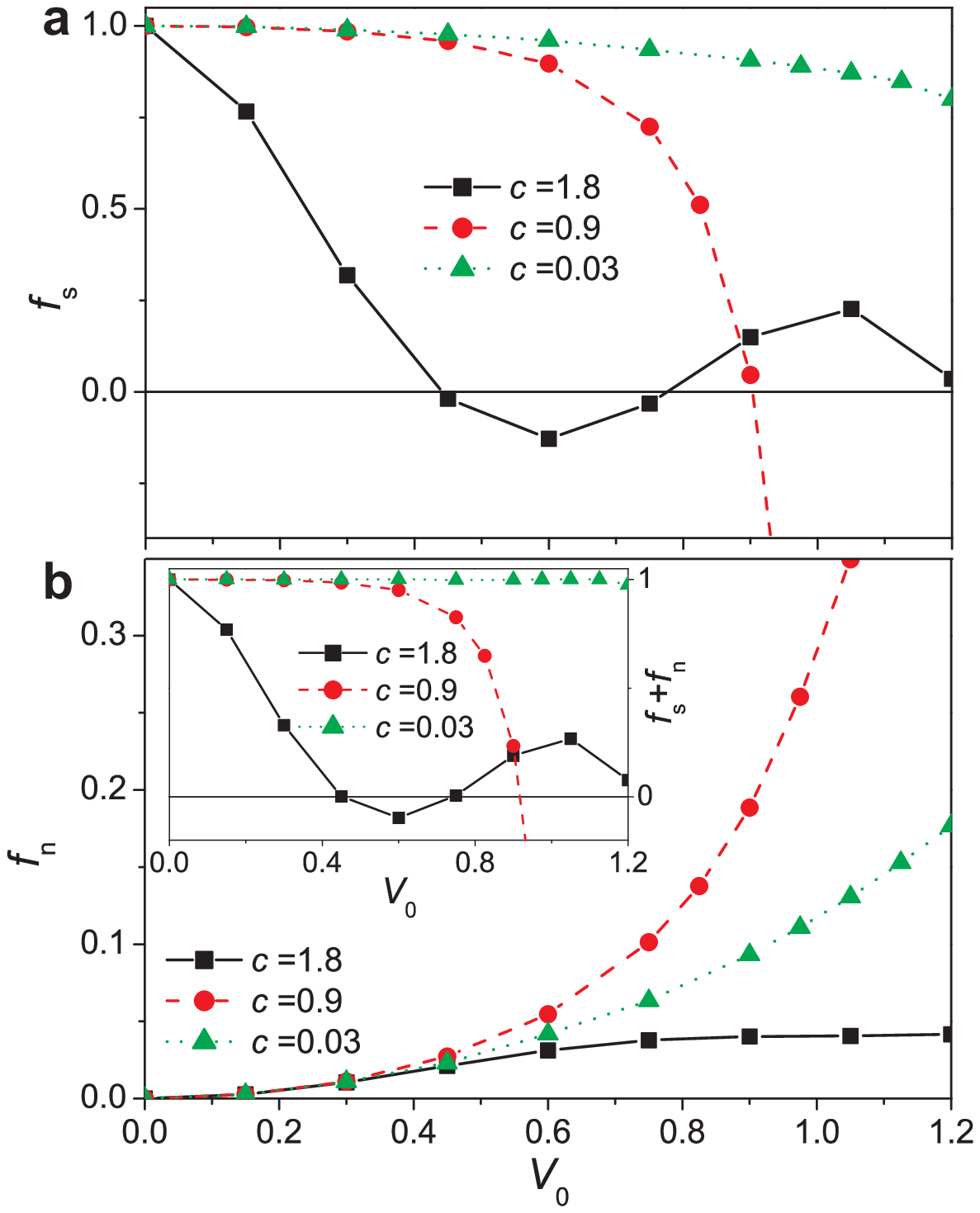}
\caption{$f_s$ (panel a) and $f_n$ (panel b) as a function of $V_0$
for $L_x = 26$, $N_x = 128$, $\nu =2$, $p = 1$, $A_x = 0.04$ and
different $c$. Inset: $f_s+f_n$ as a function of $V_0$.}
\label{Fig4}
\end{figure}

The equality $f_s+f_n=1$ is seen to be satisfied only at $V_0\to 0$.
Like in our previous example, when suppressing the pumping-loss term
(in the present case this is realised by decreasing the parameter
$c$) so that the system approaches an equilibrium regime, the
behaviour of $f_s(V_0)$ and $f_n(V_0)$ becomes more reasonable (see
Fig.~\ref{Fig4}). In particular, as seen from the inset to
Fig.~\ref{Fig4}, the range of potential strengths $V_0$, where the
condition $f_s+f_n=1$ is obeyed, increases clearly with decreasing
$c$, so that at $c=0.03$ this range includes $V_0\sim 1$.

\begin{figure}[h!]
\centering
\includegraphics*[width=1\linewidth]{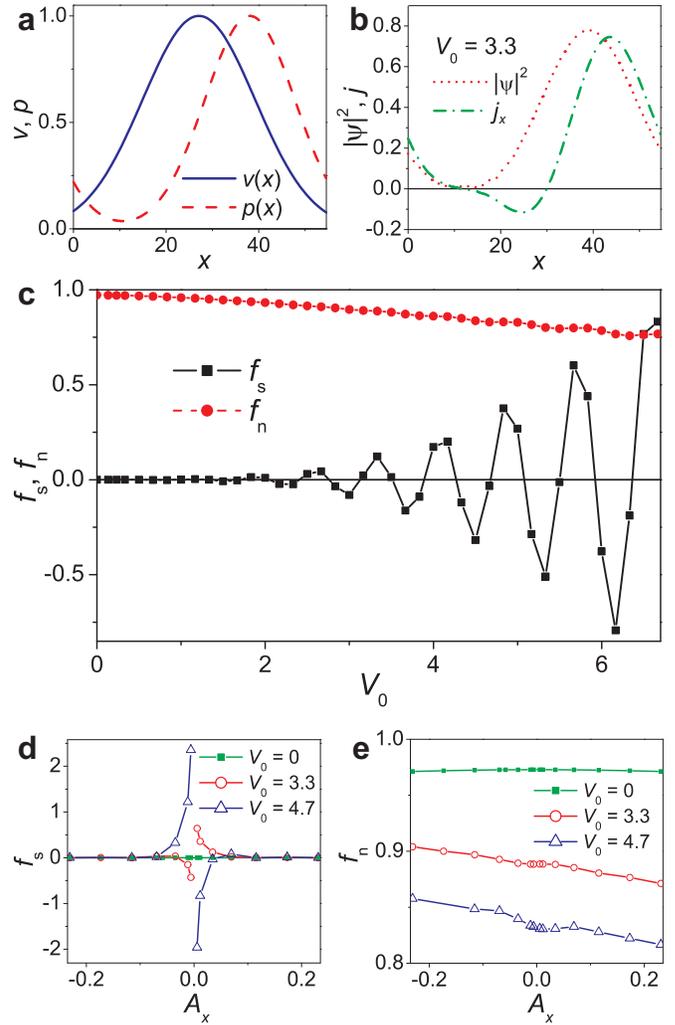}
\caption{(a) Functions $v(x)$ (solid line)  and $p(x)$ (dashed
line), which determine the spatial distributions of the static
potential and the pumping intensity, respectively, in a periodic
system with $L_x = 55$, $N_x = 128$. (b) Distributions of the
density $|\psi|^2$ (dotted line) and current density $j_x$
(dash-dotted line) at  $\nu=3$, $c = 4$, and $V_0 = 3.3$. (c) $f_s$
and $f_n$, calculated for $\nu=3$, $c = 4$, $A_x = 0.035$, as a
function of $V_0$. Panels d and e show $f_s$ and $f_n$,
respectively, as a function of $ A_x$ for $\nu=3$, $c = 4$ and three
different $V_0$. } \label{Fig5}
\end{figure}

In our last example we consider a system where a strong asymmetry is
induced by an inhomogeneous pumping with the intensity maximum
shifted with respect to the extrema of the potential $V_0v(x)$ (see
Fig.~\ref{Fig5}a). The system is uniform in the $y$ direction. As
displayed in Fig.~\ref{Fig5}b, the combined effect of the potential
barrier and inhomogeneous pumping leads to a very non-uniform
density distribution with strong currents, mainly in the positive
$x$ direction. In Fig.~\ref{Fig5}c, at relatively small $V_0$,
reasonable, nearly zero values of $f_s$ are accompanied by a
counterintuitive decrease of $f_n$ with increasing the height $V_0$
of the potential barrier. At larger $V_0$, both $f_s(V_0)$ and
$f_n(V_0)$ demonstrate some oscillatory behaviour, which can hardly
have any physical meaning. One can also notice that the condition
$f_s+f_n=1$ is not fully satisfied even at $V_0=0$.
As seen from Fig.~\ref{Fig5}d, at sufficiently large $V_0$, the
calculated function of $f_s(A_x)$ can have a pronounced
discontinuity at $|A_x|\to 0$, which is similar to that shown in the
inset to Fig.~\ref{Fig1}c for the case of a random static potential.
The origin of this discontinuity can be qualitatively understood by
associating the presence of a nonzero average current in the system
at $A_x=0$ with an effective vector potential $A_{0x}\neq 0$. Then,
assuming like before that the chemical potential of the superfluid
is proportional to the vector potential squared, $\omega(A)=\alpha
A^2$, one obtains from Eq.~(\ref{fs}) the expression $f_s=
(1+2A_{0x}/A_x)\alpha$, which obviously diverges at $|A_x|\to 0$.
The existence of a nonzero average current in the absence of an
externally applied vector potential seems to be responsible also for
the asymmetry of the calculated function $f_n(A_x)$ at large $V_0$,
both in the present example (see Fig.~\ref{Fig5}e) and in the case
of a random potential (see the inset in Fig.~\ref{Fig1}c).

\section{Conclusions}

Summarising our numerical findings, we can conclude that the
definitions \eqref{fn} and \eqref{fs} lose their physical
interpretation in terms of normal and superfluid fractions
in the presence of driving and decay. For what concerns the
definition of the superfluid fraction \eqref{fs}, this may
not be that surprising, since the stationary state is no
longer the one with minimal (free) energy. For the case of a
disordered potential, however, the superfluid fraction shows
an expected behavior as a function of disorder strength,
starting from one at zero disorder and falling to zero for
stronger disorder.

For other potential profiles however, such as a single potential
dip, the equilibrium definition of the superfluid fraction becomes
problematic. It becomes negative for certain values of the
potential. This means the frequency of the condensate shows a
decrease instead of an increase under the application of a phase
twist in the boundary condition. In equilibrium, this is forbidden,
since time reversal invariance guarantees that the condensate wave
function in the ground state is real. Out of equilibrium, where time
reversal invariance is broken, the steady state also has currents
for periodic boundary conditions. The interplay between this
existing current and the applied phase twist can then lead to a
lower frequency. In the case of a spatially inhomogeneous pump, the
behavior of the superfluid fraction even differs more dramatically
from the equilibrium one, showing large oscillations as function of
the external potential strength (see Fig. \ref{Fig5}c).

At first, the definition of the normal fraction, based on the
current response of the system, could be thought to be more robust
with respect to the nonequilibrium condition. Our numerical analysis
however strongly contradicts this. Already in the case of the
disordered potential, where the superfluid fraction behaves reasonably, the normal fraction does not show the expected
behavior.  We also attribute the failure of the equilibrium
definition of the superfluid fraction to the presence of currents in
the steady state. These currents invalidate the physical picture that the current response is due to a limited stiffness of the condensate phase.
When currents are present in the steady state, also a redistribution
of the density will contribute to the current response. The
interplay between the density redistribution and the steady state
currents can then even lead to negative normal fraction, as we
observed in the case of the disordered potential landscape (Fig.
\ref{Fig1}c). The conclusion of our numerical findings is negative:
the equilibrium definitions for superfluid and normal fractions no
longer give meaningful results out of equilibrium in the presence of
external potentials that induce currents.

It will be of interest to include the effect of fluctuations in the
calculation of the superfluid and normal fractions. Within the
classical field formalism, this could be done in the truncated
Wigner approximation \cite{wouters2009}. Of specific interest are
the superfluid and normal fraction close to the phase transition. At
equilibrium, it is of the BKT type in two dimensions, with a
characteristic jump in the superfluid fraction. Indications that the
BKT character of the phase transition is preserved out of
equilibrium were found recently in numerical simulations in the
parametric oscillation regime \cite{dagvadorj2015}, but the behavior
of the normal and superfluid fractions has not been addressed yet.

\input{snfract1A.bbl}

%
%
%
%
%
%
%
%
%
%
%

\end{document}

%% file: snfract1A.bbl
%